\documentclass[12pt]{article}

\usepackage[totalwidth=460truept,totalheight=620truept]{geometry}
\usepackage{latexsym,graphicx,amsfonts,amssymb,amsmath,xcolor}
\usepackage[hypertexnames=false,hidelinks]{hyperref}

\global\arraycolsep=1truept

\begin{document}

\null

\vskip.0truecm

\begin{center}
{\huge \textbf{Purely Virtual Particles in}}

\vskip.6truecm

{\huge \textbf{Quantum Gravity,}}

\vskip.7truecm

{\huge \textbf{Inflationary Cosmology}}

\vskip.8truecm

{\huge \textbf{and Collider Physics}}

\vskip1truecm

\textsl{Damiano Anselmi}

\vskip.1truecm

{\small \textit{National Institute of Chemical Physics and Biophysics, R\"{a}%
vala 10, Tallinn 10143, Estonia}}

{\small \textit{Dipartimento di Fisica \textquotedblleft
E.Fermi\textquotedblright , Universit\`{a} di Pisa, Largo B.Pontecorvo 3,
56127 Pisa, Italy}}

{\small \textit{INFN, Sezione di Pisa, Largo B. Pontecorvo 3, 56127 Pisa,
Italy}}

{\small damiano.anselmi@unipi.it}

\vskip1truecm

\textbf{Abstract}
\end{center}

We review the concept of purely virtual particle and its uses in quantum
gravity, primordial cosmology and collider physics. The fake particle, or
\textquotedblleft fakeon\textquotedblright , which mediates interactions
without appearing among the incoming and outgoing states, can be introduced
by means of a new diagrammatics. The renormalization coincides with the one
of the parent Euclidean diagrammatics, while unitarity follows from spectral
optical identities, which can be derived by means of algebraic operations.
The classical limit of a theory of physical particles and fakeons is
described by an ordinary Lagrangian plus Hermitian, micro acausal and micro
nonlocal self-interactions. Quantum gravity propagates the graviton, a
massive scalar field (the inflaton) and a massive spin-2 fakeon, and leads
to a constrained primordial cosmology, which predicts the tensor-to-scalar
ratio $r$ in the window $0.4\lesssim 1000r\lesssim 3.5$. The interpretation
of inflation as a cosmic RG\ flow allows us to calculate the perturbation
spectra to high orders in the presence of the Weyl squared term. In models
of new physics beyond the standard model, fakeons evade various
phenomenological bounds, because they are less constrained than normal
particles. The resummation of self-energies reveals that it is impossible to
get too close to the fakeon peak. The related peak uncertainty, equal to the
fakeon width divided by 2, is expected to be observable.

\vfill\eject

\section{Introduction}

\label{intro}\setcounter{equation}{0}

Nature \textquotedblleft is written in that great book which ever is before
our eyes -- I mean the universe -- but we cannot understand it if we do not
first learn the language and grasp the symbols in which it is written. The
book is written in mathematical language, and the symbols are triangles,
circles and other geometrical figures...\textquotedblright\ Since Galileo's
time, the language of the book of nature has evolved considerably. For some
time, the power of infinitesimal calculus gave us the illusion of the
continuum and determinism. Then, unexpectedly, quantum mechanics turned
everything upside down, by injecting uncertainty into the laws of physics.
Mathematics successfully made room for the new concepts, but several
problems remained unresolved, or so it appeared to us. With the advent of
quantum field theory (renormalization, the challenges of perturbation theory
and the impossibility to move beyond the perturbative expansion in a
systematic way), the mathematization of physical phenomena became more
challenging.

As far as we know today, the language spoken by the elementary particles is
diagrammatic. And consequently, perturbative. Beyond that we have hints, but
no satisfactory formal setup. A nonperturbative language might even not
exist.

Strictly speaking, there is no reason why nature should be mathematizable by
one of the living species it generates around the universe. In the end, we
are just clumps of atoms and logic is a net of brain connections among
memorized, mainly acoustic perceptions, which are our words (hence \textit{%
the} word, verb, or logos), shaped by experience through repetition, custom
and mental habit (\`{a} la Hume), rather than having an existence \textit{%
per se}, although the idea of logic existing \textquotedblleft before
nature\textquotedblright\ is one of those which hardly die and periodically
come back under different spells. Actually, our size and the relative scales
involved in the phenomena of the universe suggest that the logicization of
nature is most likely impossible beyond certain limits. The question is
whether we reached ours or there is still room for improvement.

The challenge of quantum gravity inspired many to call everything into
question again (this time, for free) and suggest thoroughly new approaches
\textquotedblleft beyond\textquotedblright\ quantum field theory, despite
the lack of data pointing to such a turmoil. Probably, the underlying
assumption was, again, that logic is not just a tool, a language, but
pre-exists nature (\textit{in principio erat verbum}), so we should be able
to grasp the theory (of something, or even everything) without or with very
little experimental data, by overthinking (banking on a special connection
with divinities?) or following personal or social tastes (\textquotedblleft
string theory is so beautiful that it can only be true\textquotedblright ).

\textquotedblleft It's the diagrammatics, stupid\textquotedblright : what
if, instead, quantum gravity were just a step away from the standard model?
Just a fairly guessable missing piece of the puzzle? The diagrammatic
approach has worked very well so far for the elementary particles and the
standard model. Yet, despite decades of efforts, there is still a lot to
understand about it and the basic principles on which it is founded, which
are locality, unitarity and renormalizability. Quantum field theory never
ceases to surprise, so to speak.

A new concept in the world of diagrammatics is the concept of purely virtual
particle, which we review in this paper together with its applications.
Purely virtual particles, or fake particles, or \textquotedblleft
fakeons\textquotedblright , are \textquotedblleft non
particles\textquotedblright , or particles that have no classical limit.
That is to say, they have a purely quantum nature. Their effects reach the classical
limit as effective interactions among the physical particles. Normally, we
take for granted that everything belonging to the quantum world should be
obtained by quantizing something classical, but if we view the matter the
other way around, we can easily make room for entities that are classically
hidden and exist only at the quantum level.

The diagrammatics of fakeons \cite{diagrammarMio} is obtained from the usual
one by means of surgical operations that selectively remove degrees of
freedom at all energies and preserve the optical theorem. The only
requirement is that fake particles should be massive and non tachyonic. The
main application of fakeons is the formulation of a consistent theory of
quantum gravity \cite{LWgrav}, which is observationally testable due to its
predictions in inflationary cosmology \cite{ABP}. At the phenomenological
level, fakeons evade common constraints that limit the employment of normal
particles. Among the other things, they can be used to propose new physics
beyond the standard model \cite{Tallinn1} and solve discrepancies with data 
\cite{Tallinn2}. We stress that the fakeon diagrammatics is relatively
straightforward, to the extent that it can be implemented in software like
FeynCalc, FormCalc, LoopTools and Package-X \cite{calc} and used to work out
physical predictions. For proofs to all orders, see \cite%
{fakeons,diagrammarMio}. It is also possible \cite{LWformulation,FakeonsLW} to
avoid certain troubles of the Lee-Wick models \cite%
{leewick,LWQED,lee,nakanishi,CLOP,grinstein} by switching to theories of
particles and fakeons. Finally, the fakeon prescription can be used to give sense 
to higher-spin massive multiplets \cite{HS}.
Coupled to gravity, higher-spin massive multiples 
change the ultraviolet behavior and open the way
to asymptotic freedom \cite{HSPiva}.

The paper is organized as follows. In section \ref{introintro} we review
some key concepts concerning unitarity. In section \ref{diagrammar} we
introduce the fakeon diagrammatics. In section \ref{QG} we briefly recall
how the quantization of gravity works by means of fakeons. In section \ref{r}
we discuss the main predictions of quantum gravity with fakeons in
primordial cosmology. In section \ref{phenom} we present some ways to use
fakeons in phenomenology. In section \ref{microc} we discuss the main two
new features of the theories with fakeons: the peak uncertainty and the
violation of microcausality. Section \ref{conclusions} contains conclusions
and outlook.

\section{Particles, fakeons and ghosts}

\label{introintro}\setcounter{equation}{0}

Unitarity is the statement that the scattering matrix $S$ is unitary, $%
S^{\dagger }S=1$. Writing $S=1+iT$, it is also expressed by the optical
theorem 
\begin{equation}
iT-iT^{\dagger }+T^{\dagger }T=0,  \label{opti}
\end{equation}%
which admits a diagrammatic, off-shell version in terms of identities%
\begin{equation}
G+\bar{G}+\sum_{c}G_{c}=0  \label{cut}
\end{equation}%
among cut diagrams \cite{unitarity}. Here $G$ denotes an ordinary (uncut)
diagram and stands for $iT$, $\bar{G}$ is its complex conjugate and stands
for $-iT^{\dagger }$, while $G_{c}$ are the so-called cut diagrams, obtained
by cutting internal lines:\ they stand for $T^{\dagger }T$. The vertices and
propagators that lie to one side of the cut are the normal ones (as in $T$),
while those that lie to the other side of the cut are the complex conjugate
ones (as in $T^{\dagger }$). The cut propagators give us information about
the on-shell content of a particle. The equations (\ref{cut}) single out
certain analytic properties of the loop integrals, which encode, among the
other things, the physical processes where the virtual particles circulating
in the loops turn real, which occurs above certain thresholds. A purely
virtual particle cannot turn real, by definition, so its cut propagator must
vanish.

Denoting the space of physical states by $V$ and inserting a complete set of
orthonormal states $|n\rangle \in V$, equation (\ref{opti}) implies, in
particular, 
\begin{equation}
2\hspace{0.01in}\mathrm{Im}\langle a|T|a\rangle =\sum_{|n\rangle \in
V}|\langle n|T|a\rangle |^{2},  \label{o2}
\end{equation}%
where $|a\rangle $ $\in V$ is an arbitrary state: the total cross section
for production of all final states is proportional to the imaginary part of
the forward scattering amplitude. The simplest cutting equations are 
\begin{eqnarray}
2\hspace{0.01in}\text{Im}\left[ (-i)%
\raisebox{-1mm}{\scalebox{2}{$\rangle
\hspace{-0.075in}-\hspace{-0.07in}\langle$}}\,\right] =%
\raisebox{-1mm}{\scalebox{2}{$\rangle
\hspace{-0.075in}-\hspace{-0.14in}\slash\hspace{-0.015in}\langle$}} &=&\int 
\mathrm{d}\Pi _{f}\hspace{0.01in}\left\vert \raisebox{-1mm}{\scalebox{2}{$%
\rangle\hspace{-0.035in}-$}}\right\vert ^{2},  \label{cutd} \\
2\hspace{0.01in}\text{Im}\left[ (-i)\raisebox{-1mm}{\scalebox{2}{$-%
\hspace{-0.065in}\bigcirc\hspace{-0.065in}-$}}\right] =\hspace{0.01in}%
\raisebox{-1mm}{\scalebox{2}{$-\hspace{-0.065in}\bigcirc\hspace{-0.16in}%
\slash\hspace{0.015in}-$}} &=&\int \mathrm{d}\Pi _{f}\hspace{0.01in}%
\left\vert \raisebox{-1mm}{\scalebox{2}{$-\hspace{-0.035in}\langle$}}%
\right\vert ^{2},  \label{cutdd}
\end{eqnarray}%
where the integrals are over the phase spaces $\Pi _{f}$ of the final
states. In particular, (\ref{cutd}) implies Re[$P$]$\geqslant 0$, if $P$ is
the propagator.

Physical particles, ordinary ghosts, Lee-Wick (LW) ghosts and purely virtual
particles have propagators%
\begin{equation*}
\frac{i}{p^{2}-m^{2}+i\epsilon },\qquad -\frac{i}{p^{2}-m^{2}+i\epsilon }%
,\qquad -\frac{i}{p^{2}-m^{2}-i\epsilon },\qquad \pm \mathcal{P}\frac{i}{%
p^{2}-m^{2}},
\end{equation*}%
respectively, where $\mathcal{P}$ denotes the Cauchy principal value. They
all satisfy Re[$P$]$\geqslant 0$, except for the ordinary ghost, which
violates unitarity. The propagators of physical particles and ordinary
ghosts can be used \textquotedblleft as is\textquotedblright\ inside Feynman
diagrams, which means as they appear in the formulas just written, by
integrating on real loop energies and momenta. Instead, the propagators of
LW ghosts and purely virtual particles cannot, because the $i\epsilon $ and $%
-i\epsilon $ prescriptions cannot coexist inside Feynman diagrams without
violating unitarity, the locality and Hermiticity of counterterms and
stability \cite{Wheelerons}. These two options need suitable integration
prescriptions or, in the case of fakeons, a new diagrammatics.

The removal of degrees of freedom from the incoming and outgoing states is
consistent only if it is compatible with unitarity, in which case it is
called \textquotedblleft projection\textquotedblright\ and the reduced
action is called \textquotedblleft projected action\textquotedblright . This
means that the equation (\ref{o2}) holds in a subspace $V$ of the total
space $W$ of states one uses to build the theory. Working in an extended
space $W$ and projecting to $V$ at the end is normally useful to manipulate
simpler Feynman rules, like those of a local theory.

A well-known example of projection is the one concerning the Faddeev-Popov
ghosts and the longitudinal/temporal components of the gauge fields in gauge
theories. There, the consistency of the projection is ensured\ by the
symmetry. In the case of the LW ghosts, instead, one has to make them
unstable, to kick them out of the set of strictly asymptotic states (which
are to be taken literally at $t=\pm \infty $):\ the projection is the very
same decay of the LW\ ghosts. In the case of fakeons, the consistency is
ensured by the diagrammatics, so there is no need for giving fakeons
nonvanishing widths, dynamically or explicitly. The \textquotedblleft
width\textquotedblright\ of a purely virtual particle has a completely
different physical interpretation. It is the \textquotedblleft peak
uncertainty\textquotedblright , which measures the impossibility of
experimentally approaching the fakeon too closely. The fakeon projection is
compatible with unitarity order by order (and diagram by diagram) in the
perturbative expansion \cite{diagrammarMio}.

\section{Purely virtual particles: a new diagrammatics}

\label{diagrammar}\setcounter{equation}{0}

The simplest way to introduce fakeons is by means of the diagrammatics
developed in ref. \cite{diagrammarMio}, which is useful for physical
particles as well. It is based on the threshold decomposition of ordinary
(cut and uncut) diagrams and the suppression of all the thresholds that
involve fakeon frequencies. The fakeon procedure works with both signs in
front of the propagators (fake particles and fake ghosts), since a sign flip
can at most flip the overall signs of the identities (\ref{cut}), which
encode unitarity, thus keeping them valid. For definiteness, we concentrate
on fakeons obtained from physical particles.

At the tree level, we start from the usual Feynman prescription, decompose
the propagators by means of the identity%
\begin{equation}
\frac{i}{x+i\epsilon }=\mathcal{P}\frac{i}{x}+\pi \delta (x)  \label{iden}
\end{equation}%
and suppress all the delta functions that refer to fakeons.

Apart from some caveats, this simple recipe can be implemented to all
orders. The key ingredient is the possibility of reducing the optical
theorem to a set of purely algebraic operations and identities. In brief,
the procedure is:

--- ignore the integral on the space components of the loop momenta (which
defines the \textit{skeleton} diagram);

--- perform the integral on the loop energies by means of the residue
theorem (which can be viewed as an algebraic operation);

--- decompose the result in terms of principal values and delta functions by
means of the identity (\ref{iden});

--- organize the decomposition properly;

--- drop all the deltas that contain fakeon frequencies.

A caveat, which can be appreciated starting from the box diagram, is that
the decomposition must be properly organized, due to certain nontrivial
identities that are met along the way.

Let us illustrate the procedure on the bubble diagram, which gives the
skeleton integral%
\begin{equation*}
B^{s}=\int \frac{\mathrm{d}k^{0}}{2\pi }\prod\limits_{a=1}^{2}\frac{2\omega
_{a}}{(k-p_{a})^{2}-m_{a}^{2}+i\epsilon _{a}}=\int \frac{\mathrm{d}k^{0}}{%
2\pi }\prod\limits_{a=1}^{2}\frac{2\omega _{a}}{(k^{0}-e_{a})^{2}-\omega
_{a}^{2}+i\epsilon _{a}},
\end{equation*}%
where $k^{0}=(k^{0},\mathbf{k})$ is the loop momentum, $p_{a}^{\mu }=(e_{a},%
\mathbf{p}_{a})$ are the external momenta (one for each \textit{internal}
leg, the redundancy being useful to have more symmetric expressions) and $%
\omega _{a}=\sqrt{(\mathbf{k}-\mathbf{p}_{a})^{2}+m_{a}^{2}}$ are the
frequencies. For convenience, a product $\prod\limits_{a=1}^{2}(2\omega
_{a}) $ is inserted after dropping the integral on $\mathbf{k}$.

The residue theorem gives%
\begin{equation*}
B^{s}=-\frac{i}{e_{1}-e_{2}-\omega _{1}-\omega _{2}+i\epsilon }-\frac{i}{%
e_{2}-e_{1}-\omega _{1}-\omega _{2}+i\epsilon }.
\end{equation*}%
The threshold decomposition using identity (\ref{iden}) gives%
\begin{eqnarray}
B^{s} &=&-\mathcal{P}\frac{i}{e_{1}-e_{2}-\omega _{1}-\omega _{2}}-\pi
\delta (e_{1}-e_{2}-\omega _{1}-\omega _{2})  \notag \\
&&-\mathcal{P}\frac{i}{e_{2}-e_{1}-\omega _{1}-\omega _{2}}-\pi \delta
(e_{2}-e_{1}-\omega _{1}-\omega _{2}).  \label{Bs}
\end{eqnarray}%
Repeating the same procedure with the conjugate diagram and the cut
diagrams, we obtain the table%
\begin{equation}
\begin{tabular}{|c|c|c|c|c|}
\hline
& $\rangle \hspace{-0.18em}{\bigcirc \hspace{-0.18em}\langle }$ & $\overline{%
\rangle \hspace{-0.18em}{\bigcirc \hspace{-0.18em}\langle }}$ & $\rangle 
\hspace{-0.18em}{\bigcirc \hspace{-0.18em}\langle }\hspace{0.01in}\hspace{%
-0.45cm}\slash\hspace{0.2cm}$ & $\widetilde{\rangle \hspace{-0.18em}{%
\bigcirc \hspace{-0.18em}\langle }\hspace{0.01in}\hspace{-0.45cm}\slash%
\hspace{0.2cm}}$ \\ \hline
--- & $-i\mathcal{\hat{P}}^{12}$ & $i\mathcal{\hat{P}}^{12}$ & $0$ & $0$ \\ 
\hline
$\Delta ^{12}$ & $-1$ & $-1$ & $0$ & $2$ \\ \hline
$\Delta ^{21}$ & $-1$ & $-1$ & $2$ & $0$ \\ \hline
\end{tabular}
\label{tb1}
\end{equation}%
where%
\begin{equation}
\mathcal{P}^{ab}=\mathcal{P}\frac{1}{e_{a}-e_{b}-\omega _{a}-\omega _{b}}%
,\qquad \mathcal{\hat{P}}^{ab}=\mathcal{P}^{ab}+\mathcal{P}^{ba},\qquad
\Delta ^{ab}=\pi \delta (e_{a}-e_{b}-\omega _{a}-\omega _{b}),  \notag
\end{equation}%
and the cut diagram with a tilde is the one where the sides corresponding to 
$T$ and $T^{\dagger }$ are interchanged.

Here and below, if $C_{ij}$ denote the entries of the table, a (cut or
uncut) diagram $G_{j}$ is the $j$th column of the table ($j>1$), by which we
mean the sum%
\begin{equation}
G_{j}\equiv \sum_{i>1}C_{i1}C_{ij},  \label{columns}
\end{equation}%
where $C_{21}=1$. The spectral optical identities are the rows of the table,
by which we mean the sums%
\begin{equation}
R_{i}\equiv C_{i1}\sum_{j>1}C_{ij}=0,  \label{rows}
\end{equation}%
for $i>1$, which vanish separately. They decompose the \textquotedblleft
spectral optical theorem\textquotedblright , which is the whole table, i.e.,
the sum%
\begin{equation}
\sum_{j>1}G_{j}=\sum_{i>1}\sum_{j>1}C_{i1}C_{ij}=0  \label{spoct}
\end{equation}%
of all its entries. Finally, the optical theorem is the integral of this
identity, divided by $4\omega _{1}\omega _{2}$, over the space components $%
\mathbf{k}$ of the loop momentum, with measure $\mathrm{d}^{3}\mathbf{k}%
/(2\pi )^{3}$.

If an internal leg, say leg 1, is a fakeon, we drop the delta functions
containing its frequency from equation (\ref{Bs}) and so obtain%
\begin{equation}
B_{\text{f}}^{s}=-\mathcal{P}\frac{i}{e_{1}-e_{2}-\omega _{1}-\omega _{2}}-%
\mathcal{P}\frac{i}{e_{2}-e_{1}-\omega _{1}-\omega _{2}}.  \label{Bsf}
\end{equation}%
In table (\ref{tb1}), we drop the rows containing $\Delta ^{12}$, which
gives 
\begin{equation*}
\begin{tabular}{|c|c|c|c|c|}
\hline
& $\rangle \hspace{-0.18em}{\bigcirc \hspace{-0.18em}\langle }$ & $\overline{%
\rangle \hspace{-0.18em}{\bigcirc \hspace{-0.18em}\langle }}$ & $\rangle 
\hspace{-0.18em}{\bigcirc \hspace{-0.18em}\langle }\hspace{0.01in}\hspace{%
-0.45cm}\slash\hspace{0.2cm}$ & $\widetilde{\rangle \hspace{-0.18em}{%
\bigcirc \hspace{-0.18em}\langle }\hspace{0.01in}\hspace{-0.45cm}\slash%
\hspace{0.2cm}}$ \\ \hline
--- & $-i\mathcal{\hat{P}}^{12}$ & $i\mathcal{\hat{P}}^{12}$ & $0$ & $0$ \\ 
\hline
\end{tabular}%
\end{equation*}%
Dropping whole rows preserves the (spectral) optical theorem in an obvious
way. Moreover, the last two columns, corresponding to the cut diagrams,
disappear as well, since their surviving entries are just zeros. We can
understand their disappearance by noting that those diagrams contain a cut
fakeon leg and the cut propagator of a fakeon must vanish, because the fakeon
cannot be on shell. This leaves us with the table%
\begin{equation*}
\begin{tabular}{|c|c|c|}
\hline
& $\rangle \hspace{-0.18em}{\bigcirc \hspace{-0.18em}\langle }$ & $\overline{%
\rangle \hspace{-0.18em}{\bigcirc \hspace{-0.18em}\langle }}$ \\ \hline
--- & $-i\mathcal{\hat{P}}^{12}$ & $i\mathcal{\hat{P}}^{12}$ \\ \hline
\end{tabular}%
\end{equation*}

In the case of the skeleton triangle $T^{s}$, we can proceed similarly.
Without giving details (which can be found in ref. \cite{diagrammarMio}),
the decomposition is%
\begin{equation}
T^{s}=-i\mathcal{P}_{\text{T}}-\sum_{\text{perms}}\Delta ^{ab}\mathcal{Q}%
^{ac}+\frac{i}{2}\sum_{\text{perms}}\Delta ^{ab}(\Delta ^{ac}+\Delta ^{cb}),
\label{Tdecomp}
\end{equation}%
where%
\begin{equation*}
\mathcal{P}_{\text{T}}=\mathcal{P}^{12}\mathcal{P}^{13}+\text{cycl}%
+(e\rightarrow -e),\qquad \mathcal{Q}^{ab}=\mathcal{P}^{ab}-\mathcal{P}\frac{%
1}{e_{a}-e_{b}-\omega _{a}+\omega _{b}},
\end{equation*}%
and the sums are on $\{a,b,c\}$ equal to the permutations of 1, 2 and 3. The
conjugate diagram is $\bar{T}^{s}$ and the cut diagrams read 
\begin{equation}
T_{\text{c}}^{s}=2\Delta ^{21}(\mathcal{Q}^{23}-i\Delta ^{31}-i\Delta
^{23}),\qquad \widetilde{T_{\text{c}}^{s}}=2\Delta ^{12}(\mathcal{Q}%
^{13}+i\Delta ^{13}+i\Delta ^{32}),  \label{cutT}
\end{equation}%
plus the ones obtained by cyclically permuting 1, 2 and 3.

If the internal leg 3 is a fakeon, the rows containing $\Delta ^{13}$, $%
\Delta ^{23}$, $\Delta ^{31}$ and $\Delta ^{32}$ must be suppressed. Then
the cut diagrams containing a cut leg 3 become trivial and their columns
disappear automatically. We remain with the table 
\begin{equation}
\begin{tabular}{|c|c|c|c|c|}
\hline
& $T_{\text{f}}^{s}$ & $\bar{T}_{\text{f}}^{s}$ & $T_{\text{fc}}^{s}$ & $%
\widetilde{T_{\text{fc}}^{s}}$ \\ \hline
--- & $-i\mathcal{P}_{\text{T}}$ & $i\mathcal{P}_{\text{T}}$ & $0$ & $0$ \\ 
\hline
$\Delta ^{12}$ & $-\mathcal{Q}^{13}$ & $-\mathcal{Q}^{13}$ & $0$ & $2%
\mathcal{Q}^{13}$ \\ \hline
$\Delta ^{21}$ & $-\mathcal{Q}^{23}$ & $-\mathcal{Q}^{23}$ & $2\mathcal{Q}%
^{23}$ & $0$ \\ \hline
\end{tabular}
\label{Tf}
\end{equation}%
If two internal legs are fakeons, the last two rows disappear, which make
the last two columns disappear as well:%
\begin{equation*}
\begin{tabular}{|c|c|c|}
\hline
& $T_{\text{ff}}^{s}$ & $\bar{T}_{\text{ff}}^{s}$ \\ \hline
--- & $-i\mathcal{P}_{\text{T}}$ & $i\mathcal{P}_{\text{T}}$ \\ \hline
\end{tabular}%
\end{equation*}

Other examples (triangle with \textquotedblleft diagonal\textquotedblright ,
box, box with diagonal, pentagon, hexagon, etc.) and the proof to all orders
can be found in ref. \cite{diagrammarMio}. The threshold decomposition and
the fakeon diagrammatics are compatible with gauge invariance and general
covariance, through the WTST\ identities \cite{WTST}. Indeed, the WTST\
identities are algebraic relations among the integrands of certain diagrams,
so the decomposition and the fakeon projection go through them
straightforwardly. Gauge independence is preserved as well, since the
thresholds associated with the gauge-trivial modes depend on the
gauge-fixing parameters and cannot interfere with the other
(physical/fakeon) thresholds, which are gauge invariant and gauge
independent.

\section{Quantum gravity}

\label{QG}\setcounter{equation}{0}

Quantum gravity with fakeons propagates the graviton, a scalar field $\phi $
of mass $m_{\phi }$ (the inflaton) and a spin 2 field $\chi _{\mu \nu }$ of
mass $m_{\chi }$. It is formulated starting from the classical action%
\begin{equation}
S_{\text{QG}}=-\frac{1}{16\pi G}\int \mathrm{d}^{4}x\sqrt{-g}\left( 2\Lambda
+R+\frac{\lambda }{2m_{\chi }^{2}}C_{\mu \nu \rho \sigma }C^{\mu \nu \rho
\sigma }-\frac{R^{2}}{6m_{\phi }^{2}}\right) ,  \label{sqg}
\end{equation}%
where $C_{\mu \nu \rho \sigma }$ is the Weyl tensor, $G$ is the Newton
constant, $\Lambda $ is the cosmological constant and $\lambda ={m_{\chi
}^{2}(3m_{\phi }^{2}+4\Lambda )}/(m_{\phi }^{2}(3m_{\chi }^{2}-2\Lambda ))$
is a parameter very close to 1. The theory is renormalizable by power
counting \cite{stelle}, since the renormalizability of a theory with fakeons
coincides with the one of the Euclidean parent theory.

The three fields can be made explicit by eliminating the higher derivatives
as shown in \cite{Absograv}. In particular, the action $S_{\chi }(g,\phi
,\chi )$ of $\chi _{\mu \nu }$ is the sum%
\begin{equation}
S_{\chi }(g,\phi ,\chi )=-\frac{\lambda }{8\pi G}S_{\text{PF}}(g,\chi
)+S_{\chi \text{int}}(g,\phi ,\chi )  \label{scc}
\end{equation}%
of a term proportional to the nonminimally coupled covariantized Pauli-Fierz
action%
\begin{eqnarray}
S_{\text{PF}}(g,\chi ) &=&\frac{1}{2}\int \mathrm{d}^{4}x\sqrt{-g}\left[
D_{\rho }\chi _{\mu \nu }D^{\rho }\chi ^{\mu \nu }-D_{\rho }\chi D^{\rho
}\chi +2D_{\mu }\chi ^{\mu \nu }D_{\nu }\chi -2D_{\mu }\chi ^{\rho \nu
}D_{\rho }\chi _{\nu }^{\mu }\right.  \notag \\
&&\left. -m_{\chi }^{2}(\chi _{\mu \nu }\chi ^{\mu \nu }-\chi ^{2})+R^{\mu
\nu }(\chi \chi _{\mu \nu }-2\chi _{\mu \rho }\chi _{\nu }^{\rho })\right]
\label{SPF}
\end{eqnarray}%
plus further interactions $S_{\chi \text{int}}(g,\phi ,\chi )$, where $\chi=g^{\mu\nu}\chi_{\mu\nu}$
is the trace of $\chi_{\mu\nu}$.

Since $\Lambda $ is much smaller than $m_{\chi }^{2}$, $\lambda $ is
positive, so the $\chi _{\mu \nu }$ kinetic term has the wrong sign. This is
the reason why $\chi _{\mu \nu }$ must be quantized as a fakeon. Then $\chi
_{\mu \nu }$ is purely virtual and does not belong to the sets of incoming
and outgoing states.

It is convenient to postpone the fakeon projection to the very end, to deal
with local diagrammatic rules. An early projection forces us to work with
rather involved nonlocal vertices. This situation is similar to the one of
gauge theories, where it is preferable to work with the local diagrammatic
rules of a gauge-fixed action propagating gauge-trivial modes and
Faddeev-Popov ghosts and remove them only at the very end.

The projection must also be performed classically. In this sense, the action
(\ref{sqg}) does not describe the true classical limit, because it is
unprojected. The true classical action, which is useful to study primordial
cosmology, is obtained by \textquotedblleft classicizing\textquotedblright\
quantum gravity \cite{classicalQG} and collects the tree diagrams that only
have physical particles on the external legs.

\section{Inflationary cosmology from quantum gravity}

\label{r}\setcounter{equation}{0}

Quantum gravity with fakeons can be used to study primordial cosmology and
work out predictions that could even be tested within our lifetime. For this
purpose, it is convenient the consider the action (\ref{sqg}) at $\Lambda =0$%
, make the inflaton field $\phi $ explicit through a field redefinition and
keep the fakeon $\chi _{\mu \nu }$ implicit. We obtain the equivalent action%
\begin{equation}
S_{\text{QG}}=-\frac{1}{16\pi G}\int \mathrm{d}^{4}x\sqrt{-g}\left( R+\frac{1%
}{2m_{\chi }^{2}}C_{\mu \nu \rho \sigma }C^{\mu \nu \rho \sigma }\right) +%
\frac{1}{2}\int \mathrm{d}^{4}x\sqrt{-g}\left( D_{\mu }\phi D^{\mu }\phi
-2V(\phi )\right) ,  \label{sqgeq}
\end{equation}%
where 
\begin{equation}
V(\phi )=\frac{3m_{\phi }^{2}}{32\pi G}\left( 1-\mathrm{e}^{\phi \sqrt{16\pi
G/3}}\right) ^{2}  \label{staropote}
\end{equation}%
is the Starobinsky potential.

As said, the classical limit is not described by either (\ref{sqg}) or (\ref%
{sqgeq}), which are unprojected. The classicization is nontrivial when the
metric is expanded around curved backgrounds rather than flat space.
Nevertheless, if the background is the FLRW metric, the degrees of freedom
decouple from one another at the quadratic level in the de Sitter limit \cite%
{ABP}. Thanks to this fact, the fakeon projection can be perturbatively
obtained from the flat-space one.

It can be shown that this procedure works under the consistency condition $%
m_{\chi }>m_{\phi }/4$ \cite{ABP}. This lower bound on the mass of the
fakeon $\chi _{\mu \nu }$ with respect to the mass of the inflaton $\phi $
is crucial for the prediction on the tensor-to-scalar ratio $r$, which is
determined within less than an order of magnitude, even before knowing the
actual value of $m_{\chi }$ \cite{ABP}.

Note that the theory does not predict other degrees of freedom besides the
curvature perturbation $\mathcal{R}$ and the tensor perturbations, when the
matter sector is switched off. The fakeon projection eliminates the
possibility of having additional scalar and tensor perturbations, as well as
vector perturbations.

\subsection{Cosmic RG\ flow}

Parametrizing the background metric as $g_{\mu \nu }=$diag$%
(1,-a^{2},-a^{2},-a^{2})$, the Friedmann equations and the $\phi $ equation
read%
\begin{equation}
\dot{H}=-4\pi G\dot{\phi}^{2},\qquad H^{2}=\frac{4\pi G}{3}\left( \dot{\phi}%
^{2}+2V(\phi )\right) ,\qquad \ddot{\phi}+3H\dot{\phi}=-V^{\prime }(\phi ),
\label{frie}
\end{equation}%
where $H=\dot{a}/a$ is the Hubble parameter. For the purposes of this paper,
we can assume $\dot{\phi}>0$. Defining the conformal time%
\begin{equation}
\tau =-\int_{t}^{+\infty }\frac{\mathrm{d}t^{\prime }}{a(t^{\prime })}
\label{tau}
\end{equation}%
and the \textquotedblleft coupling\textquotedblright 
\begin{equation}
\alpha =\sqrt{\frac{4\pi G}{3}}\frac{\dot{\phi}}{H}=\sqrt{-\frac{\dot{H}}{%
3H^{2}}},  \label{alf}
\end{equation}%
it is easy to show that $\alpha $ satisfies an equation of the form $\beta
_{\alpha }=\mathrm{d}\alpha /\mathrm{d\ln }|\tau |$, where $\beta _{\alpha }$
is a function of $\alpha $ that can be worked out to arbitrarily high orders
in $\alpha $: 
\begin{equation}
\beta _{\alpha }=-2\alpha ^{2}\left[ 1+\frac{5}{6}\alpha +\frac{25}{9}\alpha
^{2}+\frac{383}{27}\alpha ^{3}+\mathcal{O}(\alpha ^{4})\right] .
\label{beta}
\end{equation}

The interpretation of inflation as a \textquotedblleft cosmic\ RG
flow\textquotedblright , $\beta _{\alpha }$ being the beta function, is
predicated on the possibility of viewing the perturbation spectra $\mathcal{P%
}_{T}$ and $\mathcal{P}_{\mathcal{R}}$ of the tensor and scalar fluctuations
as correlation functions that satisfy RG evolution equations of the
Callan-Symanzik type, in the superhorizon limit \cite{CMBrunning}.

Let us introduce the running coupling $\alpha (x)$, which is the solution of%
\begin{equation*}
\ln \frac{\tau }{\tau ^{\prime }}=\int_{\alpha (-\tau ^{\prime })}^{\alpha
(-\tau )}\frac{\mathrm{d}\alpha }{\beta _{\alpha }(\alpha )}.
\end{equation*}%
For brevity, $\alpha $ will stand for $\alpha (-\tau )$ and $\alpha _{k}$
for $\alpha (1/k)$, where $k$ is just a constant for now:%
\begin{equation*}
\ln (-k\tau )=\int_{\alpha _{k}}^{\alpha }\frac{\mathrm{d}\alpha ^{\prime }}{%
\beta _{\alpha }(\alpha ^{\prime })}.
\end{equation*}%
At the leading-log level, the running coupling reads%
\begin{equation}
\alpha =\frac{\alpha _{k}}{1+2\alpha _{k}\ln (-k\tau )}.  \label{arun}
\end{equation}%
Its expression to the next-to-next-to leading log (NNLL) order can be found
in \cite{CMBrunning}.

Viewing the spectra as functions of $\tau $ and $\alpha $, their RG
evolution equations are%
\begin{equation}
\frac{\mathrm{d}\mathcal{P}}{\mathrm{d}\ln |\tau |}=\left( \frac{\partial }{%
\partial \ln |\tau |}+\beta _{\alpha }(\alpha )\frac{\partial }{\partial
\alpha }\right) \mathcal{P}=0.  \label{RG}
\end{equation}%
Viewing them as functions of $\alpha $ and $\alpha _{k}$, the dependence on $%
\alpha $ actually drops out and the spectra depend on the momentum $k$ only
through the running coupling $\alpha _{k}$: 
\begin{equation}
\mathcal{P}=\mathcal{\tilde{P}}(\alpha _{k}),\qquad \frac{\mathrm{d}\mathcal{%
\tilde{P}}(\alpha _{k})}{\mathrm{d}\ln k}=-\beta _{\alpha }(\alpha _{k})%
\frac{\mathrm{d}\mathcal{\tilde{P}}(\alpha _{k})}{\mathrm{d}\alpha _{k}}.
\label{noalfa}
\end{equation}
Finally, viewing the spectra as functions of $k/k_{\ast }$ and $\alpha
_{\ast }=\alpha (1/k_{\ast })$, where $k_{\ast }$ is the pivot scale and $%
\alpha _{\ast }$ is the \textquotedblleft pivot coupling\textquotedblright ,
they satisfy%
\begin{equation}
\left( \frac{\partial }{\partial \ln k}+\beta _{\alpha }(\alpha _{\ast })%
\frac{\partial }{\partial \alpha _{\ast }}\right) \mathcal{P}(k/k_{\ast
},\alpha _{\ast })=0.  \label{RGeq}
\end{equation}

The correspondence between the cosmic RG flow and the one of quantum field
theory is summarized in table \ref{tableI}. 
\begin{table}[t]
\begin{center}
\begin{tabular}{rcl}
QFT\ RG\ flow &  & Cosmic RG\ flow \\ \hline
RG\ flow & \multicolumn{1}{l}{$\leftrightarrow $} & slow roll \\ 
couplings $\alpha $, $\lambda $ ... & \multicolumn{1}{l}{$\leftrightarrow $}
& slow-roll parameters $\epsilon $, $\delta $ ... \\ 
beta functions & \multicolumn{1}{l}{$\leftrightarrow $} & equations of the
background metric \\ 
sliding scale $\mu $ & \multicolumn{1}{l}{$\leftrightarrow $} & conformal
time $\tau $ (or $\eta =-k\tau $) \\ 
correlation functions & \multicolumn{1}{l}{$\leftrightarrow $} & 
perturbation spectra \\ 
Callan-Symanzik equation & \multicolumn{1}{l}{$\leftrightarrow $} & RG
equation at superhorizon scales \\ 
RG\ invariance & \multicolumn{1}{l}{$\leftrightarrow $} & conservation on
superhorizon scales \\ 
asymptotic freedom & \multicolumn{1}{l}{$\leftrightarrow $} & de Sitter
limit in the infinite past \\ 
subtraction scheme & \multicolumn{1}{l}{$\leftrightarrow $} & Einstein
frame, Jordan frame, etc. \\ 
dimensional transmutation & \multicolumn{1}{l}{$\rightarrow $} & $\tau $
drops out from the spectra, \textquotedblleft replaced\textquotedblright\ by 
$k$ \\ 
running coupling & \multicolumn{1}{l}{$\rightarrow $} & ok \\ 
resummation of leading logs & \multicolumn{1}{l}{$\rightarrow $} & ok \\ 
anomalous dimensions & \multicolumn{1}{l}{$\rightarrow $} & 0%
\end{tabular}%
\end{center}
\caption{Correspondence between QFT RG flow and cosmic RG flow}
\label{tableI}
\end{table}


\subsection{Spectra}

In high-energy physics, a low-energy effective theory is good enough to make
predictions about low energies. In cosmology, it is not so: we must properly
treat the high-energy (sub-horizon) limit, even if our purpose is just to
make predictions about the low-energy (super-horizon) limit. This is a
highly nontrivial problem, since the sub-horizon region is experimentally
and observationally inaccessible. We can say something reasonable about it
only if the system reduces to one we have experience of around us. This is
where fakeons play a crucial role in primordial cosmology.

If $\chi _{\mu \nu }$ is quantized by means of the Feynman prescription
instead of the fakeon one, the theory has ghosts and so violates unitarity 
\cite{stelle}. From the point of view of primordial cosmology, the problem
of ghosts shows up as follows.

On a nontrivial background, the study of the metric fluctuations reduces, in
the end, to the problem of harmonic oscillators with time-dependent
frequencies. We need to provide a proper quantization condition to study
such a system. Normally, the Bunch-Davies vacuum condition \cite{BD} is
chosen, which does refer to the sub-horizon limit of the theory, where the
problem can be handled because the frequencies of the oscillators becomes
time independent. If ghosts are present, no matter how heavy they are, they
do not disappear at high energies, but just become massless. A condition
like the Bunch-Davies one on ghost oscillators is not robust, even if their
frequencies are constant, because we do not have examples of elementary systems
of that type that can justify it.

The situation changes in the theory with fakeons. We must ensure that the\
fakeons are indeed fake at all scales, including the sub-horizon ones. In
the low energy regime fakeons disappear for free, because they massive, but
in the opposite limit the consistency of the fakeon projection and in
particular its classicization \cite{classicalQG} on a curved background,
requires that we impose a condition, which is the bound $m_{\chi }>m_{\phi
}/4$ of ref. \cite{ABP}. In the end, this condition turns out to be rather
powerful, because it gives constrained predictions, even if $m_{\chi }$ is
still unknown. We see that fakeons provide a second reason, besides the
Bunch-Davies vacuum condition, why we must properly treat the high energies
to make predictions about the low energies in primordial cosmology.

The spectra of the theory with ghosts are studied in\ \cite{salvio} and the
comparison with those of the theory with fakeons, which we report below, can
be found in \cite{ABP}.

We briefly describe the strategy of the calculation in the theory with
fakeons. First, the metric is expanded as%
\begin{eqnarray}
g_{\mu \nu } &=&\text{diag}(1,-a^{2},-a^{2},-a^{2})-2a^{2}\left( u\delta
_{\mu }^{1}\delta _{\nu }^{1}-u\delta _{\mu }^{2}\delta _{\nu }^{2}+v\delta
_{\mu }^{1}\delta _{\nu }^{2}+v\delta _{\mu }^{2}\delta _{\nu }^{1}\right) ,
\notag \\
&&+2\text{diag}(\Phi ,a^{2}\Psi ,a^{2}\Psi ,a^{2}\Psi )-\delta _{\mu
}^{0}\delta _{\nu }^{i}\partial _{i}B-\delta _{\mu }^{i}\delta _{\nu
}^{0}\partial _{i}B  \label{mets}
\end{eqnarray}%
in the comoving gauge, where $u=u(t,z)$ and $v=v(t,z)$ are the tensor
fluctuations and $\Psi $, $B$ are the other scalar fluctuations. The $\phi $
fluctuation $\delta \phi $ is set to zero by a gauge choice, so the
curvature perturbation $\mathcal{R}$\ coincides with $\Psi $. For reviews on
the parametrizations of the fluctuations, see \cite{baumann}. Second, the
action (\ref{sqgeq}) is expanded to the quadratic order in the fluctuations.
Third, the higher derivatives are eliminated by introducing extra fields.
Forth, the new Lagrangian is diagonalized in the de Sitter limit $\alpha =0$%
. Fifth, the fakeon projection is performed, which means that the fakeon
fields are integrated out by means of (the classical limit of) the fakeon
prescription. Sixth, a number of field redefinitions and time
reparametrizations are applied to cast the action into the standard
Mukhanov-Sasaki form. Seventh, the equations of motion are solved with the
Bunch-Davies vacuum condition. Eighth, all the transformations are undone,
to get to the desired two-point functions and the spectra of the
fluctuations in the super-horizon limit. For details, see \cite{CMBrunning}.

Thanks to the RG techniques presented above, \textquotedblleft RG
improved\textquotedblright\ tensor and scalar power spectra $\mathcal{P}_{T}$
and $\mathcal{P}_{\mathcal{R}}$ can be worked out to high orders. This means
that, although $\mathcal{P}_{T}$ and $\mathcal{P}_{\mathcal{R}}$ are
expanded in powers of $\alpha _{\ast }$, the product $\alpha _{\ast }\ln
(k/k_{\ast })$ is considered of order zero and treated exactly. The results
to the NNLL order are 
\begin{eqnarray}
\mathcal{P}_{T}\left( k\right) &=&\frac{4m_{\phi }^{2}\zeta G}{\pi }\left[
1-3\zeta \alpha _{k}\left( 1+2\alpha _{k}\gamma _{M}+4\gamma _{M}^{2}\alpha
_{k}^{2}-\frac{\pi ^{2}\alpha _{k}^{2}}{3}\right) +\frac{\zeta ^{2}\alpha
_{k}^{2}}{8}(94+11\xi )\right.  \notag \\
&&\left. +3\gamma _{M}\zeta ^{2}\alpha _{k}^{3}(14+\xi )-\frac{\zeta
^{3}\alpha _{k}^{3}}{12}(614+191\xi +23\xi ^{2})+\mathcal{O}(\alpha _{k}^{4})%
\right] ,  \label{ptF} \\
\mathcal{P}_{\mathcal{R}}(k) &=&\frac{Gm_{\phi }^{2}}{12\pi \alpha _{k}^{2}}%
\left[ 1+(5-4\gamma _{M})\alpha _{k}+\left( 4\gamma _{M}^{2}-\frac{40}{3}%
\gamma _{M}+\frac{7}{3}\pi ^{2}-\frac{67}{12}-\frac{\xi }{2}F_{\text{s}}(\xi
)\right) \alpha _{k}^{2}+\mathcal{O}(\alpha _{k}^{3})\right]  \notag
\end{eqnarray}%
where%
\begin{eqnarray*}
\xi &=&\frac{m_{\phi }^{2}}{m_{\chi }^{2}},\qquad \zeta =\left( 1+\frac{\xi 
}{2}\right) ^{-1},\qquad \gamma _{M}=\gamma _{E}+\ln 2, \\
F_{\text{s}}(\xi ) &=&1+\frac{\xi }{4}+\frac{\xi ^{2}}{8}+\frac{\xi ^{3}}{8}+%
\frac{7\xi ^{4}}{32}+\frac{19}{32}\xi ^{5}+\frac{295}{128}\xi ^{6}+\frac{1549%
}{128}\xi ^{7}+\frac{42271}{512}\xi ^{8}+\mathcal{O}(\xi ^{9})
\end{eqnarray*}%
$\gamma _{E}$ being the Euler-Mascheroni constant. While $\mathcal{P}_{T}$
is exact in $\xi $, so far the NNLL contribution to $\mathcal{P}_{\mathcal{R}%
}$ has been determined only as an asymptotic expansion in powers of $\xi $.

\subsection{Predictions}

A number of other quantities can be calculated from the spectra, such as the
\textquotedblleft dynamical\textquotedblright\ tensor-to-scalar ratio 
\begin{equation}
r(k)=\frac{\mathcal{P}_{T}(k)}{\mathcal{P}_{\mathcal{R}}(k)}  \label{dynr}
\end{equation}%
the tilts%
\begin{equation*}
n_{T}=-\beta _{\alpha }(\alpha _{k})\frac{\partial \ln \mathcal{P}_{T}}{%
\partial \alpha _{k}},\qquad n_{\mathcal{R}}-1=-\beta _{\alpha }(\alpha _{k})%
\frac{\partial \ln \mathcal{P}_{\mathcal{R}}}{\partial \alpha _{k}},
\end{equation*}%
and the running coefficients%
\begin{equation*}
\frac{\mathrm{d}^{n}n_{T}}{\mathrm{d}\ln k\hspace{0.01in}^{n}}=\left( -\beta
_{\alpha }(\alpha _{k})\frac{\partial }{\partial \alpha _{k}}\right)
^{n}n_{T},\qquad \frac{\mathrm{d}^{n}n_{\mathcal{R}}}{\mathrm{d}\ln k\hspace{%
0.01in}^{n}}=\left( -\beta _{\alpha }(\alpha _{k})\frac{\partial }{\partial
\alpha _{k}}\right) ^{n}n_{\mathcal{R}}.
\end{equation*}%
Using (\ref{ptF}), we find, for example,%
\begin{eqnarray}
n_{T} &=&-6\left[ 1+4\gamma _{M}\alpha _{k}+(12\gamma _{M}^{2}-\pi
^{2})\alpha _{k}^{2}\right] \zeta \alpha _{k}^{2}+\left[ 24+3\xi +4(31+2\xi
)\gamma _{M}\alpha _{k}\right] \zeta ^{2}\alpha _{k}^{3}  \notag \\
&&\qquad \qquad -\frac{1}{8}(1136+566\xi +107\xi ^{2})\zeta ^{3}\alpha
_{k}^{4}+\mathcal{O}(\alpha _{k}^{5}),  \label{tilts} \\
n_{\mathcal{R}}-1 &=&-4\alpha _{k}+\frac{4\alpha _{k}^{2}}{3}(5-6\gamma
_{M})-\frac{2\alpha _{k}^{3}}{9}(338-90\gamma _{M}+72\gamma _{M}^{2}-42\pi
^{2}+9\xi F_{\text{s}})+\mathcal{O}(\alpha _{k}^{4}).\qquad  \notag
\end{eqnarray}%
The first two corrections to the usual relation $r+8n_{T}\simeq 0$ are%
\begin{equation}
r+8n_{T}=-192\zeta \alpha _{k}^{3}+8(202\zeta +65\xi \zeta -144\gamma
_{M}-8\pi ^{2}+3\xi F_{\text{s}})\zeta \alpha _{k}^{4}+\mathcal{O}(\alpha
_{k}^{5}).  \label{ratiodevia}
\end{equation}

We discuss the validity of the predictions by expressing the results in
terms of a pivot\ scale $k_{\ast }$ and evolving $\alpha (1/k)$ from $%
k_{\ast }$ to $k$ by means of the RG evolution equations. The spectra become
functions of $\ln (k_{\ast }/k)$ and the pivot coupling $\alpha _{\ast
}\equiv \alpha (1/k_{\ast })$. With $k_{\ast }=0.05$ Mpc$^{-1}$ and (for
definiteness) $\xi \sim F_{\text{s}}\sim 1$, the data\ reported in \cite%
{Planck18} give $\ln (10^{10}\mathcal{P}_{\mathcal{R}}^{\ast })=3.044\pm
0.014$ and $n_{\mathcal{R}}^{\ast }=0.9649\pm 0.0042$, where the star
superscript means that the quantity is evaluated at the pivot scale. The
second formula of (\ref{ptF}) and formula (\ref{tilts}) give the values 
\begin{equation*}
\alpha _{\ast }=0.0087\pm 0.0010,\qquad m_{\phi }=(2.99\pm 0.37)\cdot 10^{13}%
\text{GeV}
\end{equation*}%
for the \textquotedblleft fine structure constant\textquotedblright\ $\alpha
_{\ast }$ and the inflaton mass, respectively. The value of $m_{\chi }$ will be known as
soon as the tensor-to-scalar ratio $r$ will be measured. The bound $m_{\chi
}>m_{\phi }/4$ gives $4\cdot 10^{-4}\lesssim r\lesssim 3.5\cdot 10^{-3}$ at
the pivot scale.

The first formula of (\ref{ptF}) predicts the tensor spectrum $\mathcal{P}%
_{T}$ with a relative theoretical error $\sim\alpha _{\ast }^{4}\sim
10^{-8}$. The relative error on the tensor tilt $n_{T}$ is $\sim\alpha _{\ast
}^{3}\sim 10^{-6}$. As far as the quantities involving the scalar
fluctuations are concerned, we have to take into account that the function $%
F_{\text{s}}(\xi )$ is only partially known. It can be shown that the
relative theoretical errors of the scalar spectrum $\mathcal{P}_{\mathcal{R}%
} $ and the scalar tilt $n_{\mathcal{R}}-1$ are around $\alpha _{\ast
}^{3}\sim 10^{-6}$ for $\xi <1/2$, $10^{-5}$ for $1/2<\xi <1$ and $10^{-4}$
for $1<\xi <16$.

If primordial cosmology turns into an arena for precision tests of quantum
gravity, the predictions might have a chance to be tested in the incoming years 
\cite{CMBStage4}.

\section{Phenomenology of fake particles}

\label{phenom}\setcounter{equation}{0}

Fakeons can be used to propose models of new physics beyond the standard
model. For example, the popular inert doublet model \cite{IDM} has rather
different phenomenological properties if the second doublet is taken to be a
fakeon \cite{Tallinn1}. Since the fake doublet avoids the $Z$-pole
constraints regardless of the chosen mass scale, there is room for new
effects below the electroweak scale. In addition, the absence of on-shell
propagation prevents fakeons from inducing missing energy signatures in
collider experiments.

Other types of standard model extensions by means of fakeons predict
measurable interactions at energy scales that are usually precluded. For
example, the interactions between a fake scalar doublet and the muon can
explain discrepancies concerning the measurement of the muon anomalous
magnetic moment \cite{Tallinn2}. The experimental results can be matched for
fakeon masses below the electroweak scale without contradicting precision
data and collider bounds on new light degrees of freedom.

An important topic for the phenomenology of particle physics is the
treatment of dressed propagators. Since a fakeon appears to have a sort of
\textquotedblleft mass\textquotedblright\ and a sort of \textquotedblleft
width\textquotedblright , but it is not a particle, we should provide
physical meanings for such two quantities. In the next section we explain
that the mass is the scale of the violation of microcausality. The width,
instead, has a thoroughly new interpretation.

The resummation of self-energy diagrams into dressed propagators in the case
of purely virtual particles reveals some unexpected facts, which, in turn,
highlight nontrivial properties of long-lived unstable particles. We
summarize here the main points, the details being available in ref. \cite%
{FSelfK}.

We factor out the normalization factor $Z$ of the propagator. We also
include the corrections $\Delta m$ to the mass $m$ into $m$ itself by
default. This way, we can focus our attention on the width $\Gamma $, since $%
Z$ and $\Delta m$ do not play crucial roles. The formally resummed dressed
propagators of physical particles $\varphi $, fake particles $\chi $ and
ghosts $\phi $ then read, around the peaks,
\begin{eqnarray}
\hat{P}_{\varphi } &\simeq &\frac{i}{p^{2}-m^{2}+i(\epsilon +m\Gamma )}%
,\qquad \hat{P}_{\chi }\simeq \frac{i(p^{2}-m^{2})}{%
(p^{2}-m^{2})(p^{2}-m^{2}+im\Gamma )+\epsilon ^{2}},  \notag \\
\hat{P}_{\phi } &\simeq &-\frac{i}{p^{2}-m^{2}+i(\epsilon -m\Gamma )},
\label{p2}
\end{eqnarray}%
respectively. It is easy to show that they differ by infinitely many contact
terms, which do not admit well-defined sums, such as%
\begin{equation}
\Delta _{\hat{\Gamma}}(x)\equiv \sum_{n=0}^{\infty }\frac{(-\hat{\Gamma}%
^{2})^{n}}{(2n)!}\delta ^{(2n)}(x),  \label{dgn}
\end{equation}%
where $x\equiv (p^{2}-m^{2})/m^{2}$ and $\hat{\Gamma}=\Gamma /m$ ($\Gamma
\geqslant 0$). Specifically,%
\begin{equation*}
\left. \text{Im}[im^{2}(\hat{P}_{\varphi }-\hat{P}_{\phi })]\right\vert _{%
\epsilon\rightarrow 0}=2\pi \Delta _{\hat{\Gamma}}(x),\qquad \left. 
\text{Im}[im^{2}(\hat{P}_{\varphi }-\hat{P}_{\chi })]\right\vert _{
\epsilon\rightarrow 0}=\pi \Delta _{\hat{\Gamma}}(x).
\end{equation*}%
It turns out that $\Delta _{\hat{\Gamma}}(x)$ is not a well-defined
mathematical distribution. What does that mean? The problem is that the peak
region is outside the convergence domain of the geometric series and can
only be reached in the case of physical particles, from the convergence
region, by means of analyticity. In the other cases, nonperturbative effects
become important.

Not only. Ill-defined quantities also appear in the case of unstable, long-lived
physical particles, when we separate their observation from the observation
of their decay products. By the optical theorem, the imaginary part $2$Re$[%
\hat{P}_{\varphi }]$ is equal to the sum of the cross sections $\Omega
_{\varphi \hspace{0.01in}\text{particle}}$ and $\Omega _{\varphi \hspace{%
0.01in}\text{decay}}$ of the processes $e^{+}e^{-}\rightarrow \varphi $ and $%
e^{+}e^{-}\rightarrow $ decay products of $\varphi $, which can be read by
cutting the diagrams contributing to the dressed propagators. The former is
the process where the particle is physically observed before it decays (as
in the case of the muon). The latter is the process where its decay products
are observed, instead (as in the case of the $Z$ boson).

We find 
\begin{equation}
\Omega _{\varphi \hspace{0.01in}\text{particle}}\simeq \frac{{\epsilon}%
}{(p^{2}-m^{2})^{2}+({\epsilon}+m\Gamma )^{2}},\qquad \Omega _{\varphi 
\hspace{0.01in}\text{decay}}\simeq \frac{m\Gamma }{(p^{2}-m^{2})^{2}+(
\epsilon+m\Gamma )^{2}},  \label{Ms}
\end{equation}%
so the limit $\epsilon \rightarrow 0$ tells us that the muon is unobservable:%
\begin{equation}
\Omega _{\varphi \hspace{0.01in}\text{particle}}\rightarrow 0,\qquad \Omega
_{\varphi \hspace{0.01in}\text{decay}}\rightarrow \frac{m\Gamma }{%
(p^{2}-m^{2})^{2}+m^{2}\Gamma ^{2}}.  \label{limpa}
\end{equation}

This is not a surprising result, if we recall that the scattering processes
are supposed to occur between incoming\ states at $t=-\infty $ and outgoing\
states at $t=+\infty $, which makes it impossible to observe an unstable
particle. However, the observation of the muon is a fact and we should be
able to account for it.

In practical situations the scattering processes take some finite
time interval $\Delta t$, much larger than the duration $\bar{\Delta}t$ of
the interactions involved in the process, but not equal to infinity. The
prediction $\Omega _{\varphi \hspace{0.01in}\text{particle}}=0$ remains
correct when $\Delta t$ is much larger than, say, the muon lifetime $\tau
_{\mu }$, but fails for $\bar{\Delta}t\ll \Delta t\lesssim \tau _{\mu }$.

To solve the impasse, we introduce the energy resolution $\Delta E\sim 1/%
\bar{\Delta}t$. In principle, we should undertake the task of rederiving all
the basic formulas of quantum field theory for scattering processes where
incoming\ and outgoing\ states are separated by a finite $\Delta t$. The
results will depend on $\Delta E$, since $\Delta E=0$ is only compatible
with $\bar{\Delta}t=\infty $, hence $\Delta t=\infty $. A clever shortcut is
to guess how $\Delta E$ may affect the results.

Generically, we can expect that $\Delta E$ will affect the formulas more or
less everywhere. However, in most places we can neglect it, especially when
it redefines quantities that are already present (like the mass $m$). The $%
\Delta E$ dependence cannot be ignored if it affects a \textquotedblleft
zero\textquotedblright , such as the imaginary part of the denominator of
the propagator around the peak.

Thus, we assume that when $\Delta E$ is different from zero the predictions
coincide with the ones we have written above, provided we make the
replacement%
\begin{equation}
{\epsilon}\rightarrow {\epsilon}+2m\Delta E,  \label{repla}
\end{equation}%
after which we can legitimately take ${\epsilon}$ to zero. The form of
the $\Delta E$ dependence appearing here is not crucial, as long as the
correction vanishes when $\Delta E$ tends to zero. Making the replacement in
formulas (\ref{Ms}) and letting ${\epsilon}$ tend to zero, we obtain%
\begin{eqnarray}
\Omega _{\varphi \hspace{0.01in}\text{particle}} &\simeq &\frac{2m\Delta E}{%
(p^{2}-m^{2})^{2}+m^{2}(2\Delta E+\Gamma )^{2}},  \label{1} \\
\Omega _{\varphi \hspace{0.01in}\text{decay}} &\simeq &\frac{m\Gamma }{%
(p^{2}-m^{2})^{2}+m^{2}(2\Delta E+\Gamma )^{2}},  \label{2}
\end{eqnarray}%
The results show that $\Omega _{\varphi \hspace{0.01in}\text{particle}}$ is
no longer zero. Phenomenologically we may distinguish two opposite cases:

--- The case of the $Z$ boson, which is $\Delta E\ll \Gamma /2$. There, 
\begin{equation*}
\Omega _{\varphi \hspace{0.01in}\text{particle}}\simeq 0,\qquad \Omega
_{\varphi \hspace{0.01in}\text{decay}}\simeq \frac{m\Gamma }{%
(p^{2}-m^{2})^{2}+m^{2}\Gamma ^{2}},
\end{equation*}%
so we do not see the particle: we see its decay products. The results do not
depend on $\Delta E$ to the first degree of approximation.

--- The case of the muon, which is $m\gg \Delta E\gg \Gamma /2$. There,%
\begin{equation}
\Omega _{\varphi \hspace{0.01in}\text{particle}}\simeq \frac{2\Delta E}{%
(p^{2}-m^{2})^{2}+4m^{2}\Delta E^{2}}\simeq \pi \delta (p^{2}-m^{2}),\qquad
\Omega _{\varphi \hspace{0.01in}\text{decay}}\simeq 0,  \label{mu}
\end{equation}%
so we see the particle and not its decay products. Again, the results do not
depend on $\Delta E$ to the first degree of approximation.

In the intermediate situations, where $\Delta E$ and $\Gamma $ are
comparable, we see both the particle and its decay products and the results
depend on $\Delta E$.

Ultimately, this has to do with the energy-time uncertainty relation $\Delta
E\hspace{0.01in}\sim 1/\bar{\Delta}t$. Indeed, $\Delta E=0$ implies an
infinite time uncertainty, during which every unstable particle has enough
time to decay before being observed. An infinite
amount of time is required to determine an energy with absolute precision,
and such an amount of time is available only for stable particles. It is
impossible to observe an unstable particle with infinite resolving power on
its energy.

However, quantum field theory is not quantum mechanics, where wave functions
allow us to keep time, coordinates, energy and momenta, and their
uncertainty relations, under a satisfactory control.\ In quantum field
theory, as it is usually formulated, we renounce any determination of time
and coordinates and tacitly assume infinite resolving powers on energy and
momenta. This means that we have a worse control on the built-in uncertainty
relations. It may occur that we unawaredly try and calculate something that
is impossible to calculate, because it violates such relations, as in the
case of $\Omega _{\varphi \hspace{0.01in}\text{particle}}$ with no $\Delta E$.
The theory cannot return a meaningful result there, otherwise it would be in
contradiction with the premises it is built on. Not unexpectedly, we find
mathematical problems in the forms of ill-defined distributions, which may
appear term by term or in the resummations.

In the case of fakeons something similar happens, but more invasively, since
analyticity is less powerful there. Making the replacement ${\epsilon}%
\rightarrow m\Delta E$ (with a different factor with respect to (\ref{repla}%
), for convenience), the convergence region of $\hat{P}_{\chi }$ is
delimited by the condition%
\begin{equation*}
\frac{m\Gamma |p^{2}-m^{2}|}{(p^{2}-m^{2})^{2}+m^{2}\Delta E^{2}}<1,
\end{equation*}%
which holds for every $p$ if and only if%
\begin{equation}
\Delta E>\frac{\Gamma }{2}.  \label{indeto}
\end{equation}%
With the conventions just chosen, this bound coincides with the one of
physical particles. The difference is that in the case of physical particles
we can cross the obstacle by means of analyticity (unless we separate the
observation of the particle from the observation of its decay products, as said).
Instead, we cannot cross it in the case of purely virtual particles, because
the fakeon prescription is not analytic.

Ghosts exhibit somewhat similar features, in this respect, but we do not
discuss them here.

It is conceivable that (\ref{indeto}) encodes a new type of uncertainty
relation, a \textquotedblleft peak uncertainty\textquotedblright , which
expresses the impossibility of approaching the fakeon too closely, given its
nature of particle that cannot be brought to reality. It also gives a
meaning to the fakeon width, while the fakeon mass\ codifies the violation
of microcausality/microlocality.

These properties suggest that certain processes may involve nonperturbative
aspects. A way to avoid them is by restricting the invariant masses $M=\sqrt{%
p^{2}}$ of the sets of external states mediated by fakeons by means of the
conditions%
\begin{equation}
|M^{2}-m^{2}|>m\Gamma .  \label{conda}
\end{equation}%
So doing, we keep the processes far enough from the regions of the fakeon
peaks, which allows us to take $\Delta E$ to zero. Under these assumptions,
we can make predictions about scattering processes at arbitrarily high
energies.

However, conditions like (\ref{conda}) do not allow us to sum over the whole
phase spaces of the final states, because such a sum includes contributions
from the regions of the fakeon peaks. For that purpose, we may propose
effective formulas for the complete dressed propagators, argued from the
general properties of fakeons. An example is%
\begin{equation}
\hat{P}_{\chi }=\frac{i(p^{2}-m^{2})}{(p^{2}-m^{2})(p^{2}-m^{2}+im\Gamma
)+\gamma ^{2}m^{2+2\delta }\Gamma ^{2-2\delta }},  \label{dressedf}
\end{equation}%
where $\gamma $ and $\delta $ are constants, satisfying $\gamma >0$, $%
0<\delta <1$. This formula can be obtained by choosing 
\begin{equation}
\Delta E=\gamma \Gamma \left( \frac{m}{\Gamma }\right) ^{\delta },
\label{putative}
\end{equation}%
which fulfills (\ref{indeto}) in the classical limit $\Gamma \rightarrow 0$,
where (\ref{dressedf}) correctly tends to the principal value of $%
i/(p^{2}-m^{2})$. An expression like (\ref{putative}) could be originated by
nonperturbative effects or describe the impact of the experimental setup.

If some relatively light fakeon exists in nature, it should be possible to
detect the peak uncertainty experimentally. Instead of seeing a resonance,
as we expect for a normal particle, we should see a bump, or a smeared peak,
with a shape that might even depend on the experimental setup in a way that
could be difficult, or impossible, to predict.

\section{Peak uncertainty and micro acausality}

\label{microc}\setcounter{equation}{0}

A violation of microcausality, with typical scale equal to the fakeon mass,
is associated with the intrinsic nonlocal nature of the fakeon projection.
Consider the toy model described by the Lagrangian%
\begin{equation*}
\mathcal{L}(x,Q,t)=\frac{m}{2}\dot{x}^{2}-m\dot{x}\dot{Q}+\frac{mM^{2}}{2}%
Q^{2}+xF_{\text{ext}}(t),
\end{equation*}%
where $x$ is the coordinate of a physical particle of mass $m$, $Q$ is the
one of a purely virtual particle of mass $M$ and $F_{\text{ext}}(t)$ is a
time-dependent external force. The equations of motion give 
\begin{equation*}
\ddot{x}=-M^{2}Q,\qquad \ddot{Q}+M^{2}Q=-\frac{1}{m}F_{\text{ext}}(t).
\end{equation*}%
The solution of the $Q$ equation, which reads 
\begin{equation*}
mQ=-\mathcal{P}\frac{1}{\frac{\mathrm{d}^{2}}{\mathrm{d}t^{2}}+M^{2}}F_{%
\text{ext}}(t)=-\frac{1}{2M}\int_{-\infty }^{\infty }\mathrm{d}u\hspace{0in}%
\hspace{0.01in}F_{\text{ext}}(t-u)\sin \left( M|u|\right) ,
\end{equation*}%
is given by the fakeon prescription. The equation of motion for $x$ then
reads%
\begin{equation}
m\ddot{x}=\frac{M}{2}\int_{-\infty }^{\infty }\mathrm{d}u\hspace{0in}\hspace{%
0.01in}F_{\text{ext}}(t-u)\sin \left( M|u|\right) .  \label{equation}
\end{equation}%
We see that the integral appearing on the right-hand side receives
contributions from the external force in the past and in the future. Due to
the oscillating behavior of $(M/2)\sin (M|u|)$, the amount of future
effectively contributing is 
\begin{equation}
\left\vert \Delta u\right\vert \simeq \frac{1}{M}  \label{uncecaus}
\end{equation}%
and disappears for $M\rightarrow \infty $, since $\lim_{M\rightarrow \infty
}(M/2)\sin (M|u|)=\delta (u)$. Thus, (\ref{uncecaus}) implies that we cannot
make predictions for time intervals shorter than $\tau $. In principle, we
could check (\ref{equation}) a posteriori, if we manage to measure $x(t)$
and $F_{\text{ext}}(t)$ independently.

This example shows that the violation of microcausality, being encoded in
the fakeon mass, does not need a nonvanishing width and survives the
classical limit. The peak uncertainty, instead, is encoded in the radiative
corrections that give $\Gamma $, so it disappears in the classical limit.
This does not prevent us, though, from making predictions about processes
occurring at higher energies. Finally, the violation of microcausality is
always present, while it is possible to have no peak uncertainty (\ref%
{indeto}), as in the models of ref. \cite{Tallinn1}, where fakeons have
identically vanishing widths due to a $\mathbb{Z}_{2}$ symmetry.

\section{Conclusions}

\label{conclusions}\setcounter{equation}{0}

Purely virtual particles have a variety of applications, which range from
collider physics, to quantum gravity and primordial cosmology. Fakeons
mediate interactions without appearing among the incoming and outgoing
states. Their consistency with unitarity can be proved by means of algebraic
spectral optical identities. The renormalization of a theory with fakeons
coincides with the one of the parent Euclidean theory. Its classical limit
is described by an ordinary Lagrangian plus Hermitian, microscopically
acausal and nonlocal self-interactions among the physical particles.

Quantum gravity with fakeons propagates the graviton, the inflaton and a
massive spin-2 fakeon. It can be coupled straightforwardly to the standard
model. Its classicization leads to a constrained primordial cosmology,
which predicts the tensor-to-scalar ratio $r$ in the window $0.4\lesssim
1000r\lesssim 3.5$. The interpretation of inflation as a cosmic RG\ flow
allows us to calculate the perturbation spectra up to high orders.

Fakeons evade various
phenomenological constraints that apply to physical particles. It is impossible to get too
close to the fakeon peak, because of a peak uncertainty, equal to the
fakeon width divided by 2, which is expected to be observable. Instead, the fakeon mass is
associated to the scale of the violation of microcausality.

In conclusion, the fakeon diagrammatics gives quantum field theory a chance
to surpass its own limitations and offer solutions to long-standing
problems, without leaving the realm of perturbation theory and without
advocating leaps of faith or uncertain approaches, such as string theory 
\cite{string}, loop quantum gravity \cite{loop}, holography and the AdS/CFT\
correspondence \cite{ads}. The way paved by purely virtual particles tops
the competitors in calculability, predictivity and falsifiability. For
example, the sharp predictions about inflationary cosmology leave little
room for artificial adjustments, in the case of discrepancies with data.
Instead, the main weakness of string theory is its lack of predictivity,
because of the landscape of 10$^{500}$ or so false vacua \cite{landscape}.
Loop quantum gravity is extremely challenging from the mathematical point of
view, when, in contrast, the fakeon diagrammatics is a relatively simple
extension of the usual diagrammatics of physical particles. The AdS/CFT
correspondence does have a quantum field theoretical side, but it is a
strongly coupled one, which leads to use nonperturbative methods, mostly
based on conjectures. A separate discussion applies to the idea of
asymptotic safety \cite{asafety}, which is purely quantum field theoretical.
Nevertheless, it also requires nonperturbative methods, to deal with the
interacting ultraviolet fixed points.

\vskip 2 truecm \noindent {\large \textbf{Acknowledgments}}

\vskip .5 truecm

This work was supported in part by the European Regional Development Fund through the CoE program grant TK133 and the Estonian Research Council grant PRG803.

\end{document}